\documentclass[a4paper,fleqn,usenatbib]{mnras}
\usepackage[T1]{fontenc}
\usepackage{ae,aecompl}
\usepackage{graphicx}	
\usepackage{amsmath}	
\usepackage{amssymb}
\usepackage{color}

\title[Origin of GRB variabilities]{GRB variabilities and following gravitational waves induced by gravitational instability in NDAFs}

\author[N. Shahamat et al.]{Narjes Shahamat,$^{1}$\thanks{n.shahamat.d@gmail.com} Shaharam Abbassi$^{1, 2}$\thanks{abbassi@um.ac.ir} and Tong Liu$^{3}$\thanks{tongliu@xmu.edu.cn}
\\
$^{1}$Department of Physics, School of Science, Ferdowsi University of Mashhad, Mashhad, PO Box 91775-1436, Iran\\
$^{2}$Research Center for Modeling and Computational Sciences, Ferdowsi University of Mashhad, Iran\\
$^{3}$Department of Astronomy, Xiamen University, Xiamen, Fujian 361005, China
}

\date{Accepted XXX. Received YYY; in original form ZZZ}

\pubyear{2021}

\begin{document}
\label{firstpage}
\pagerange{\pageref{firstpage}--\pageref{lastpage}}
\maketitle

\begin{abstract}
The present work proposes a new formalism for the inner regions of a neutrino-dominated accretion flows (NDAFs) by considering the self-gravity, where the neutrino opacity is high enough to make neutrinos trapped becoming a dominant factor in the transportation of energy and angular momentum over the magneto rotational instability. We investigate the possibility of gravitational instability and fragmentation to model the highly variable structure of the prompt emission in gamma-ray bursts (GRBs). The results lead us to introduce the gravitational instability, in these inner regions, as a source of a new viscosity which is of the same functional form as that of the $\beta$-prescription of viscosity. Such a consideration brings about fragmentation in the unstable inner disk. In addition, we find the consequent clumpy structure of this area capable to account for the temporal variability of GRB's light curve, especially for the lower choices of the parameter $\beta$, $\sim 10^{-5}$. Finally, we predict the formation of gravitational waves through the migration of fragments before being tidally disrupted. These waves appear to be detectable via a range of current and future detectors from LIGO to Cosmic Explorer.
\end{abstract}

\begin{keywords}
accretion, accretion disks -- black hole physics -- gamma-ray bursts -- gravitational waves -- neutrinos
\end{keywords}



\section{Introduction}
\label{sec1}

Gamma-ray burst (GRB) is an immediate release of energy, $\sim 10^{51}-10^{54} ~\rm erg$, which lasts from less than $2~\rm s$, for short-duration GRBs (SGRBs), up to hundreds of seconds in case of long-duration GRBs \citep[LGRBs, for reviews, see][]{2004RvMP.. 76..1143,2015PhR.. 561.. 1}. The latter's origin is mainly attributed to the core collapse of massive stars \citep[e.g.][]{1993ApJ..405.. 273}, while the former is thought to emanate from the coalescence of neutron star (NS) binary or NS-black hole (BH) mergers \citep[e.g.][]{1989ApJ.. 336.. 360,1992ApJ.. 395L.. 83}. All these scenarios can lead to a hyperaccreting stellar-mass BH with a high mass accretion rate of about $0.001-10~M_{\odot} ~\rm s^{-1}$, which is hot and dense enough to be cooled via neutrino emission. This hyperaccretion disk is called neutrino-dominated accretion flow \citep[NDAF, see e.g.][for a review]{2017NewAR.. 79.. 1}, that is considered to be one of the plausible candidates for GRB central engines. Nevertheless, some may ascribe the LGRB origin, regarding its long time activity, to the BH life-time spin \citep[e.g.][]{2001ApJ..552..L31,2015..451..L26}. In particular, \citet{2001ApJ..552..L31} proposed that LGRBs emanate from the rapidly spinning BHs with the suspended accretion due to the large-scale magnetic torques, while SGRBs can be attributed to the hyperaccretion slowly spinning BHs.

Concerning the remarkable temporal variability of the GRB's light curve, several scenarios have been proposed. This variability has primarily led to the development of the internal shock model, in which the Lorentz factor of the jet varies with time, and consequently the generated luminosity becomes highly variable \citep[e.g.][]{1993MNRAS.. 263.. 861,1994ApJ.. 422.. 248}. Another approach introduced the dissipation process related to the turbulence or magnetic reconnection in the jet, as a responsible for these rapid fluctuations in prompt emissions \citep[e.g.][]{2009MNRAS.. 394.. L117}. From a similar point of view there are those who argue that the temporal variability may have something to do with the turbulent behavior in the emission mechanism of GRBs \citep[e.g.][]{1998ApJL..508..L25,2000ApJ..535..158,2014ApJ..786..146,2018Adv.. Sp.. Res.. 62..191}. On the other hand, some authors proposed that this temporal variation could be a clue as to the instabilities occurred in the hyperaccretion disk (as the GRB's central engine). For instance, based on the detailed treatment of the chemical equilibrium in the gas species, \citet{2007ApJ..664..1011} argued that the NDAF can be viscously and thermally unstable at an extremely high accretion rate. Moreover, \citet{2009ApJ..700..1970} proposed that the NDAF could be viscously unstable at a more moderate accretion rate when certain magnetic mechanism, such as the magnetic coupling between the plumping region and the disk, is considered. Regarding the possibility of the viscous instability, \citet{2012MNRAS..419..713}  found that the thermal equilibrium solutions for the convective hyperaccretion disk, in case of very low accretion rates, would also have a viscously unstable branch. This might lead to the sporadical mass accretion onto the BH resulting in a highly variable light curve. For the rather higher accretion rates ($\sim 0.003-0.01 ~ M_{\odot}~\rm s^{-1}$), however, \citet{2013ApJ.. 777(1).. L15} found a new branch in the thermal equilibrium curves, in which an accretion flow is viscously unstable (but thermally stable) because of efficient neutrino cooling. At even larger accretion rates ($\gtrsim 0.1 ~ M_{\odot}~\rm s^{-1}$), \citet{2007ApJ...663..437} introduced a new mechanism that might cause instabilities in hyperaccretion disk resulting in the oscillations observed in the GRB's light curve. The idea is mainly based on the fact that NDAFs become optically thick to neutrinos in the inner disk. More precisely, \citet{2002ApJ.. 579.. 706} showed that the neutrino opaque region exists inside a radius $\sim 6-40~R_{g}$ for $\dot{M}$ in the range of $0.1-10~M_{\odot}~\rm s^{-1}$, where $R_{g}=2GM_{\rm BH}/c^2$ is the Schwarzchild radius. They argue that for $\dot{M}\gtrsim 1 ~M_{\odot}~\rm s^{-1}$, neutrinos are sufficiently trapped so that flow is advection dominated in this realm. Being inspired by this idea, \citet{2007ApJ...663..437} investigated this region to probe for the effectiveness of the magnetorotational instability (MRI) and found that the energy and momentum transport by neutrinos (i.e., neutrino viscosity) could suppress the growth of the MRI significantly, when the magnetic field strength $B \lesssim 10^{14}~\rm G$ is considered. Having said that, MRI can drive active magnetohydrodynamic turbulence in the outer neutrino-transparent region regardless of the field strength. This gives birth to an accumulation of the baryon matter into the inner dead zone, where the MRI grows inactively and the accretion of matter is suppressed. When the dead zone achieves a large amount of mass and becomes gravitationally unstable, the intense mass accretion onto the BH is triggered through the gravitational torque, episodically. This process can account for the short-term variability in the prompt emission of GRBs.

Concerning the gravitational instability, another approach to treat this variability in both prompt emission and late time flaring activity can be introduced as the clumpy structure of GRB's central engine. Recently, \citet{2020ApJL..896L..38C} presented a model to deal with the variable structure of SGRB's spectrum. Their study reveals that the tails generated from compact object mergers can be gravitationally unstable, leading to the creation of small scale fragments under their own self-gravity, and consequently, to the variability in fallback rate. This can account for the temporal behavior of the GRB's light curve in prompt emission as well as the late time flaring activity (a late time variability in the GRB's light curve that occurs $100-1000~\rm s$ after the prompt emission), depending on how rapidly the instability occures. Additionally, a similar clumpy structure in the outer hyperaccretion disk has been proposed to be responsible for the late time flaring activity of the central engine \citep[e.g.,][]{2006ApJ..636..L29,2017MNRAS..464..4399,2020ApJ..888..64}.

On the other hand, the impacts of self-gravity have been investigated in the highly dense structure of NDAFs. The feature that might affect this outer disk's clumpy configuration \citep[e.g.,][]{2014ApJ...791..69,2017ApJ...845..64}. Similarly, one might think of a clumpy structure in the inner regions of a self-gravitating NDAF to model the oscillations of the prompt emissions' light curve. However, this is not a possible scenario for the inner disk when the common formalism of the NDAF is applied \citep[as studied by][]{2014ApJ...791..69,2017ApJ...845..64}. Having said that, considering the trapped neutrinos in the inner regions of NDAFs (that are assumed to be optically thick to neutrinos \citep{2002ApJ.. 579.. 706} as a dominant factor in the energy and momentum transport over MRI \citep{2007ApJ...663..437,2007ApJ...655..447}, can reduce the accretion rate effectively. This might increase the density of matter inside the NDAF as pointed out by \citet{2007ApJ...663..437}. Therefore, a self-gravitating formalism for this neutrino opaque region might provide us with an enhanced possibility for the gravitational instability, and probably, fragmentation in this inner zone.

In the hope of developing a possible scenario, we aim to model the temporal variability of the prompt emission via a self-gravitating prescription of NDAF in which neutrino viscosity acts efficiently through the inner regions instead of MRI, and consequently no $\alpha$-viscosity is considered. In other words, the advection is taken to be the dominant cooling mechanism. Probing the gravitational instability and the possibility of fragmentation in this inner region will provide us with a vision of how successful this formalism can model the temporal variability of GRB's light curve. However, as can be seen in the following sections, the presence of self-gravity only results in the instability and no fragmentation occurs through this approach, since the disk is cooling down slowly (the local cooling time is larger than the dynamical time scale of the disk). In this situation one can consider the gravitational instability as an alternative that facilitates the energy and angular momentum transport, which means another source of viscosity can be implemented \citep[see e.g.,][]{2005MNRAS..358..1489,2011ARAA..49..195,2016ApJ..824..91}. We will demonstrate that such an additional viscosity term can guarantee the fragmentation in this area as it improves the cooling rate.

This paper is organized as follows. In section \ref{sec2} our formalism for the inner regions of a self-gravitating NDAF, regarding the neutrino viscosity as a dominant factor in energy and momentum transportation, is presented. The possibility of gravitational instability and fragmentation is probed in section \ref{sec2.1}. To make fragmentation plausible, in section \ref{sec3}, we provide a new prescription of viscosity, i.e. Beta-Neutrino viscosity (a simple combination of $\beta$-viscosity term and its neutrino counterpart), identifying the gravitational instability as a source of turbulence inside the neutrino opaque region. A comparison with observational evidences is presented in section \ref{sec3.1}. Moreover, to make some predictions we study the possibility for gravitational waves (GWs) in section \ref{sec4}, in the case that fragments migrate prior to tidal disruption. Finally, section \ref{sec5} sums up the present work, in addition to highlighting our main results and providing some discussions.

\section{Physical Model in Neutrino Viscosity Prescription}
\label{sec2}

To estimate neutrino viscosity we need neutrino depth in both absorption and scattering processes. Regarding the Urca process as the absorption mechanism which contributes most with respect to the other mechanisms, such as the inverse of pair annihilation, bremsstrahlung and plasmon decay \citep[e.g.][]{2002ApJ.. 577.. 311,2007ApJ...661.1025L}, the absorptive neutrino depth is described as follows \citep{2007ApJ...663..437,2002ApJ.. 579.. 706}
\begin{equation}
\tau_{a}\approx 4.5\times 10^{-39}T^{2}X_{\rm nuc}\rho H,\nonumber\
\end{equation}
where $T$, $\rho$, and $H$ are the temperature, density, and half thickness of the disk, respectively, and $X_{\rm nuc}$ is the mass fraction of free nucleons that can be considered as unity in the inner region of the disk.

Elastic scattering from nucleons gives rise to the following relation of the scattering optical depth,
\begin{equation}
\tau_{s}\approx 8.1\times 10^{-39}T^{2}\rho H,\nonumber\
\end{equation}
in which we assumes equal contributions for three types of neutrinos (i.e., electron, $\mu$, and $\tau$ neutrinos). Consequently, the total neutrino optical depth reads
\begin{equation}
\tau_{\rm tot}\approx 1.3\times 10^{-38}T^{2}\rho H.\nonumber\
\end{equation}

Introducing the avaraged mean free path of the neutrino as $\textless\lambda\textgreater=H/\tau_{tot}$, the neutrino viscosity is described by \citep{2007ApJ...663..437}
\begin{equation}
\nu = \frac{4}{15}\frac{U_{\nu}}{\rho c}\textless\lambda\textgreater\approx 5.2\times 10^{12}T^{2}\rho ^{-2},
\label{1}
\end{equation}
where $U_{\nu}=4\sigma_{B}T^{4}/c$ is the neutrino energy density \citep{1988Phys.. Rep.. 163.. 51,2007ApJ...663..437}, and $\sigma_{B}$ is the Stefan-Boltzmann constant and $c$ is the speed of light.

Concerning the disk structure, we assume a quasi-steady axisymmetric configuration for the inner neutrino opaque regions of our self-gravitating disk, with the kinematic neutrino viscosity is taken into consideration. Therefore, we have \citep{Frank2002}
\begin{equation}
\nu \Sigma =-\frac{\dot{M}}{2\pi R}\Omega (\frac{\partial \Omega}{\partial R})^{-1},
\label{2}
\end{equation}
where we ignored the angular momentum in the inner boundary of the disk. The angular momentum $\Omega$ is assumed to be Keplerian. Such an approximation is a physically reliable one in the astrophysical systems with a rather massive accretion disk around a compact object of a comparable mass with respect to the mass of the surrounding disk. However, this is not the case for extremely massive disks \citep{Philip2020}, such as those in the protostellar, protoplanetary or active galactic nuclei (AGN), in which a deviation from Keplerian motion due to self-gravity is of the order of $H/r$ \citep{2016.. ARAA.. 54.. 271}. In these cases, an increase in angular velocity \citep{1999.. AA.. 350.. 694} is expected. This impact of the self-gravity may lead the disk to be less unstable. In the case of the GRB central engine, however, we believe that the disk mass does not meet the very massive condition (as the disk mass can be regarded in a range of $\sim 0.2-5 ~M_{\odot}$, with the central BH mass of the order of $\sim 2.3-4 ~M_{\odot}$ \citep{2015ApJ.. 815.. 54}, so that the Keplerian approximation can be still an acceptable one. Through making use of the vertically averaged approximation of the density, i.e. $\rho \simeq \frac{\Sigma}{2H}$, a combination of Equations (\ref{1}) and (\ref{2}) leads to
\begin{equation}
\Sigma \frac{\dot{M}}{2\pi R}\Omega (\frac{\partial \Omega}{\partial R})^{-1}+20.8\times 10^{12}T^{2}H^{2}=0.
\label{3}
\end{equation}

In addition, considering advection as the only effective cooling mechanism in the inner disk, one can derive the temperature $T$ as a function of radius through energy balance equation $Q_{\rm vis}^{+}=Q_{\rm adv}^{-}$, with $Q_{\rm vis}^{+}=-\frac{\dot{M}\Omega^{2}}{2\pi}\frac{d\ln\Omega}{d\ln R}$ and $Q_{\rm adv}^{-}=\frac{1}{2\pi}\frac{\xi\dot{M}c_{s}^{2}}{R^{2}}$, where $\xi=3/2$ is a dimensionless quantity of the order of unity \citep[see e.g.][]{Kato2008}. $c_{s}$ is the speed of sound and to estimate it we assume the gas pressure as a dominant factor, ignoring the radiation and electron degeneracy pressure. It is identified as a good approximation in the inner neutrino opaque regions \citep{2002ApJ.. 579.. 706}. Therefore, we consider
\begin{equation}
c_{s}^{2}=\frac{k_{B}T}{m_{p}}.
\label{4}
\end{equation}
Consequently, the temperature can be derived as
\begin{equation}
 T=\frac{GM_{\rm BH}m_{p}}{k_{B}R},
\label{5}
\end{equation}
where $M_{\rm BH}$, $m_{p}$, and $k_{B}$ are the BH mass, proton mass and Boltzmann constant, respectively.

The vertical structure of this self-gravitating disk can be accounted for by the hydrostatic balance equation
\begin{equation}
2\pi G\Sigma +\Omega^{2} H-\frac{c_{s}^{2}}{H}=0,
\label{6}
\end{equation}
The first term has something to do with the disk's self-gravity and derived through the consideration of an infinite plane's gravity with a slowly varying surface density $\Sigma$ as the radius increases \citep{1978Acta Astron..28..91,2014ApJ...791..69}. This assumption appears a reasonable one when it comes to the small region of the inner disk in GRB central engines.

Then we are left with two equations, i.e.,
\begin{equation}
\frac{\dot{M}}{2\pi R}(\frac{\partial \Omega}{\partial R})^{-1}\Omega \Sigma +20.8\times 10^{12} (H\frac{m_{p}GM_{\rm BH}}{k_{B}R})^{2}=0
\label{7}
\end{equation}
and
\begin{equation}
2\pi G\Sigma H +\Omega^{2} H^{2}-\frac{GM_{\rm BH}}{R}=0.
\label{8}
\end{equation}
The solution to this set of equations provides us with the half thickness and surface density of the inner radii, in a self-gravitating NDAF.

\subsection{Gravitational Instability and Fragmentation}
\label{sec2.1}

Once $H$ and $\Sigma$ are achieved, one can probe for the instability through checking the Toomre criterion, i.e.,
\begin{equation}
Q=\frac{c_{s}\Omega}{\pi G\Sigma}<1.
\label{9}
\end{equation}

\begin{figure}
 	 	\includegraphics[width=80.mm]{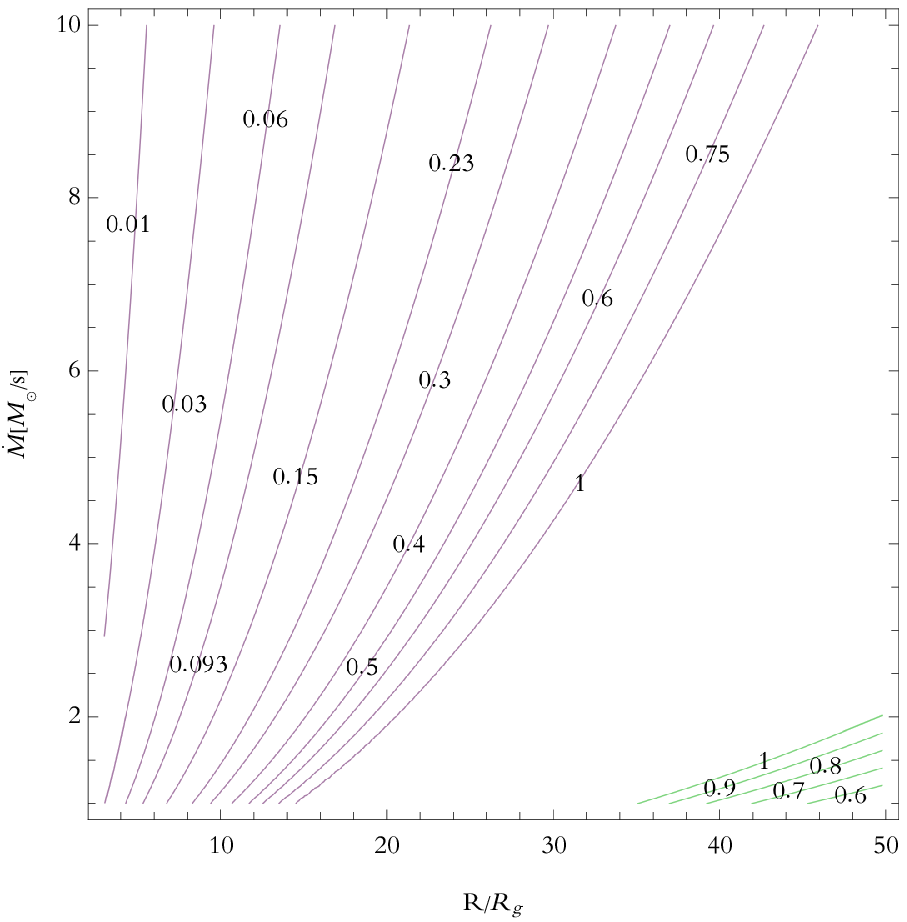}
 \includegraphics[width=80.mm]{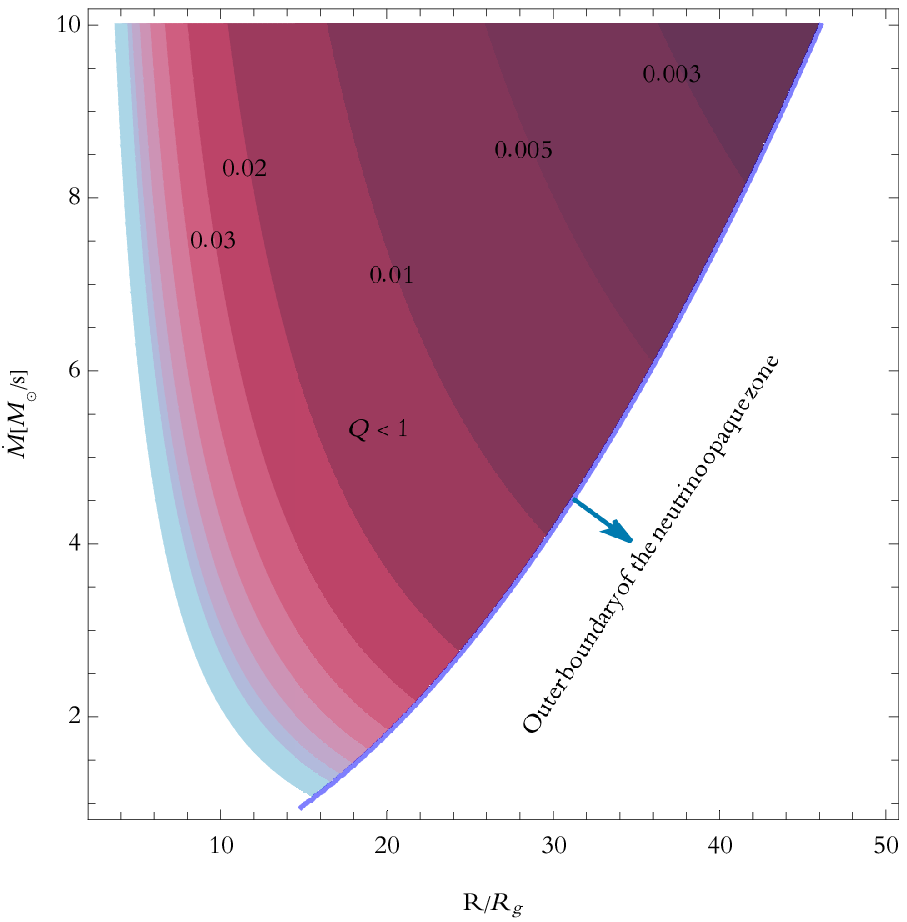}
 	\caption{\small{The top: Gravitationally
unstable regions inside the NDAF with and without self-gravity (the purple and green contours of $Q$ parameter, respectively), considering neutrino viscosity. The bottom: Contours of Toomre parameter (colored area in purple and blue) with the outer boundary of the neutrino opaque region is depicted by a thick blue line. These contours are related to the Beta-Neutrino viscosity case of a self-gravitating NDAF. }}
 	\label{fig1}
 \end{figure}

The top panel in Figure (\ref{fig1}) represents the contours of Toomre parameter in cases considering with and without self-gravity (the purple and green contours, respectively), depicted in $\dot{M}-R$ plane. The results for non-self-gravitating case is achieved through ignoring the first term in Equation (\ref{6}). It should be mentioned that we considered $M_{\rm BH}$ to be $3~M_{\odot}$. As it is obviously that, in the self-gravitating case, the upward trend of the Toomre parameter differs from its downward behavior in the case without self-gravity. In General, one can say the differential rotation works against self-gravity and has a stabilizing impact on the disk. When the self-gravity is taken into account, it affects the outer disk more effectively than the inner disk as a result of the decline in differential rotation. Thus, we think that in the outer radii the interplay between neutrino viscosity and self-gravity can lower the density rather than having an increasing impact. To elaborate more, once the self-gravity compresses these outer radii, the enhancement in the neutrino optical depth can increase the neutrino viscosity. Therefore, we are of the opinion that such an interplay can finally result in a decrease in the surface density as we go further through the disk, which is due to the inverse relation between density and neutrino viscosity ($\nu \propto \rho ^{-2}$). This can be interpreted as an upward trend of the Toomre parameter in the presence of self-gravity ($Q\propto \Sigma^{-1}$).

  \begin{figure}
 	 	\includegraphics[width=80.mm]{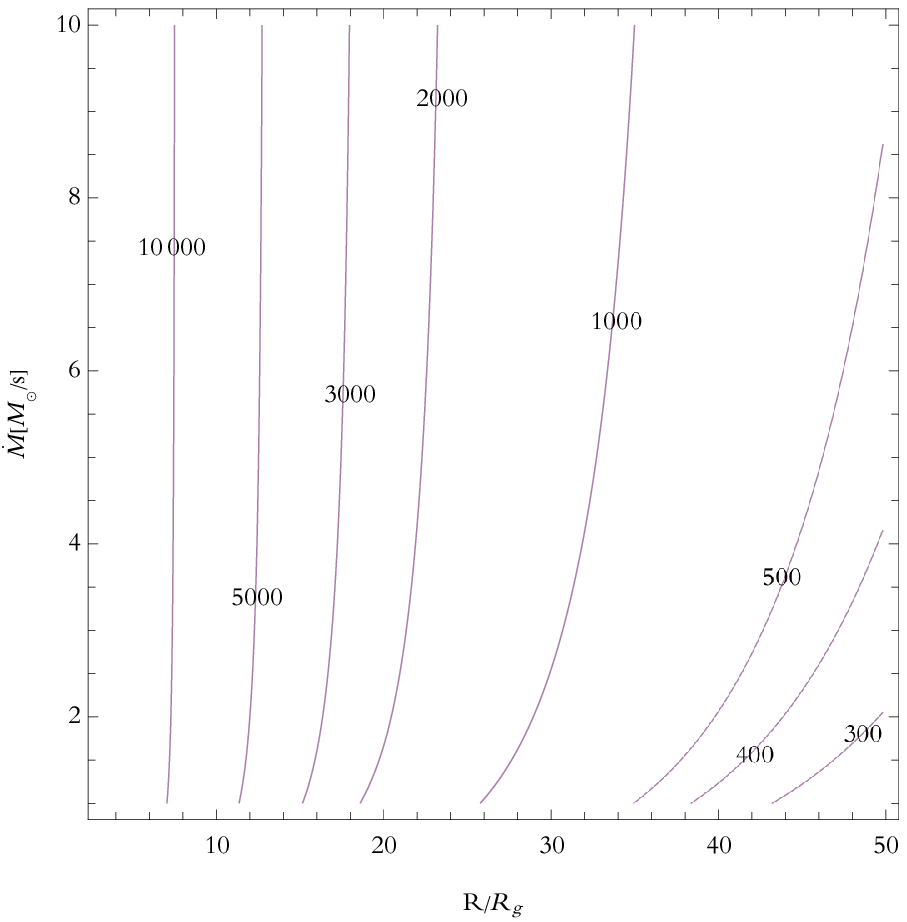}
 \includegraphics[width=80.mm]{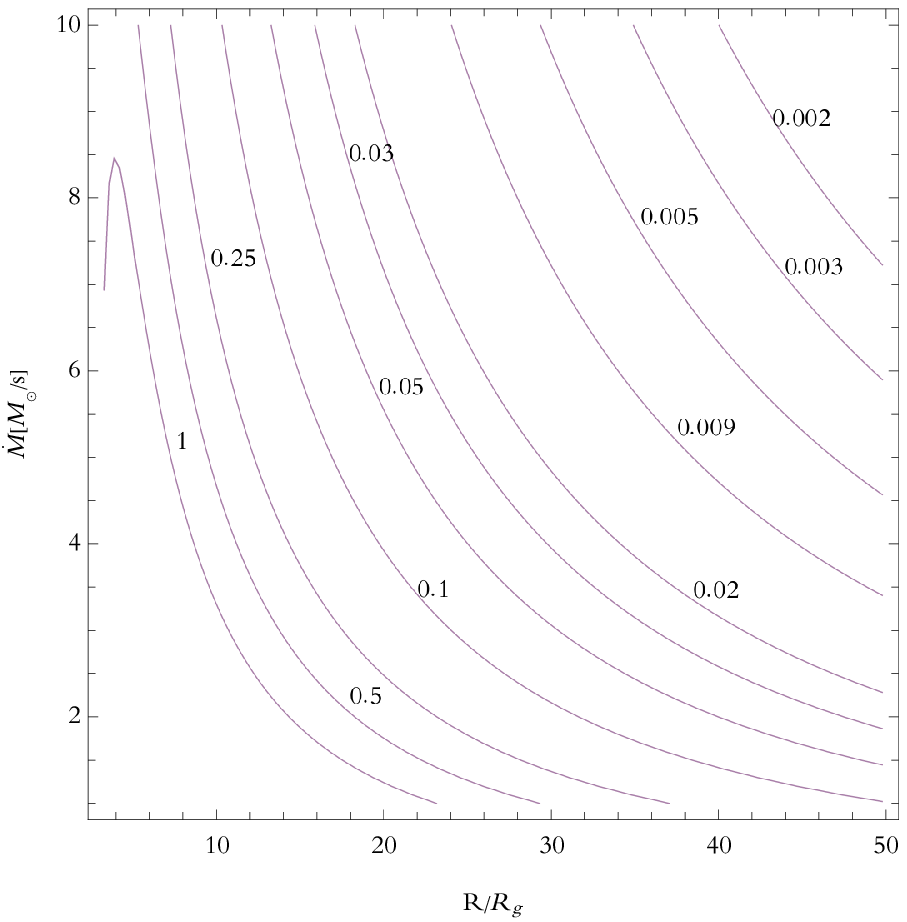}
 	\caption{\small{The top: Contours of $t_{\rm cool}/t_{\rm crit}$ in $\dot{M}-R$ plane in case of $\nu$-viscosity prescription. The bottom: Contours of $t_{\rm cool}/t_{\rm crit}$ for the case with $\beta$-viscosity. Concerning the criterion introduced in Equation (\ref{10}), the fragmentation seems possible in this scenario.}}
 	\label{fig2}
 \end{figure}

However, it is important to notice the radius inside which our formalism, with high neutrino opacity, is valid. \cite{2002ApJ.. 579.. 706} found that the optically thick region extends out to $4-5~R_{g}$ at $\dot{M}=0.1~M_{\odot}~\rm s^{-1}$, up to $30-40~R_{g}$ at $\dot{M}=10~M_{\odot}~\rm s^{-1}$. Regarding such estimates, we consider the neutrino opaque area for the NDAF to be the same as the gravitationally unstable region (the area with purple contours in the top panel of Figure (\ref{fig1})). A probably rough estimation that does not bring about a remarkable impact on our final results other than providing different numbers of data to produce a suitable fit that will appear in section \ref{sec3.1}.

As can be inferred from the green contours in this panel, there seem not to be any gravitational instability in the inner neutrino opaque regions when the self-gravity is ignored. This result reflects the fact that self-gravity matters in the inner disk using our scenario, unlike what is commonly expected from NDAF structure in which the outer disk is mainly affected by self-gravity \citep[see, e.g.][]{2017ApJ...845..64,2017NewAR.. 79.. 1}.

To probe the formation of fragments in unstable regions, the below criterion should be met \citep[e.g.,][]{2001ApJ.. 553.. 174,2005MNRAS..358..1489,2007ApJ..658..1173}
\begin{equation}
t_{\rm cool}<t_{\rm crit}\sim (1-10)~\Omega^{-1},
\label{10}
\end{equation}
where the cooling timescale is denoted by $t_{\rm cool}\approx(H/R)^{2}t_{\nu}$, with $t_{\nu}$ is obtained as below

\begin{equation}
t_{\nu}=\int_{3R_{g}}^R \! \frac{1}{v_{R}} \, \mathrm{d}r.
\label{11}
\end{equation}
The radial velocity $v_{R}$ is achievable through the continuity equation $\dot{M}=-2\pi R \Sigma v_{R}$, as well.

Considering the top panel in Figure (\ref{fig2}), which represents the contours of $t_{\rm cool}/t_{\rm crit}$ in $\dot{M}-R$ plane, one can see this criterion is not satisfied, and fragmentation is not possible in the neutrino opaque realm of a self-gravitating NDAF. However, it is argued that in such cases the disk may settle into a quasi-steady state of self-gravitating turbulence, in which an outward transport of angular momentum via gravitational torques is expected \citep{1972MNRAS..157..1,2004MNRAS..351..630,2005MNRAS..358..1489,2011ARAA..49..195,2016ApJ..824..91}. Being inspired by this idea, we are going to assume the gravitational instability as a source of turbulence, and consequently viscosity, in this unstable region of the disk. On the other hand, to introduce a functional form for this new term of viscosity we consider \citet{2006ApJ...653...L89} approach and take it to be of the same form of the $\beta$ parameterizations (i.e., $\nu=\beta R v_{\phi}$). This type of hydrodynamically driven turbulence ($\beta$-viscosity) has been proposed by \citet{1998BAAS..30..917} and \citet{1999A&A..347..734}, independently, and applied by \citet{2000A&A..357..1123} to both self-gravitating and non-self gravitating disks. They showed that this kind of viscosity yield the standard $\alpha$-disk prescription in the case of shock dissipation limited, non-selfgravitating disks \citep[for more details see also][]{2006ApJ...653...L89}. Henceforth, the viscosity in this neutrino opaque region is considered as a combination of neutrino kinematic viscosity and $\beta$ turbulent vicosity.

\section{Beta-Neutrino Viscosity Prescription}
\label{sec3}

Adding $\beta$ term to the neutrino viscosity, Equation (\ref{3}) can be rewritten in the following form,
\begin{equation}
\beta R^{2}\Omega \Sigma^{2}+\Sigma \frac{\dot{M}}{2\pi R}\Omega (\frac{\partial \Omega}{\partial R})^{-1}+20.8\times 10^{12}T^{2}H^{2}=0.
\label{12}
\end{equation}
This alters the gravitationally unstable region to what is depicted in the bottom panel of Figure (\ref{fig1}) (the colored area in purple and blue). Note that to plot the contours of $Q\textless 1$, we are bound to the neutrino opaque region as approximated in section (\ref{sec2}), so that we restricted the unstable area with a solid thick blue line as the outer boundary of the neutrino opaque zone. This panel demonstrates a different behavior of the Toomre parameter which is decreasing rather than being increasing like that of the Neutrino viscosity prescription. With the consideration of $\beta$-prescription, in addition to the neutrino viscosity, it appears this increased viscosity can lead to the lower surface density through the disk (consider Equation (\ref{2})). We believe that this can reduce the role of neutrino viscosity effectively, so that the impacts of self-gravity might not be affected by, the previously mentioned, interplay between these two factors (i.e., $\nu$-viscosity and self-gravity). Consequently, we think that the weaker stabilizing effects of the rotation accompanied by the stronger impacts of self-gravity in the outer regions can provide us with the observed downward trend of the Toomre parameter.




\begin{figure}
 	 	\includegraphics[width=75.mm]{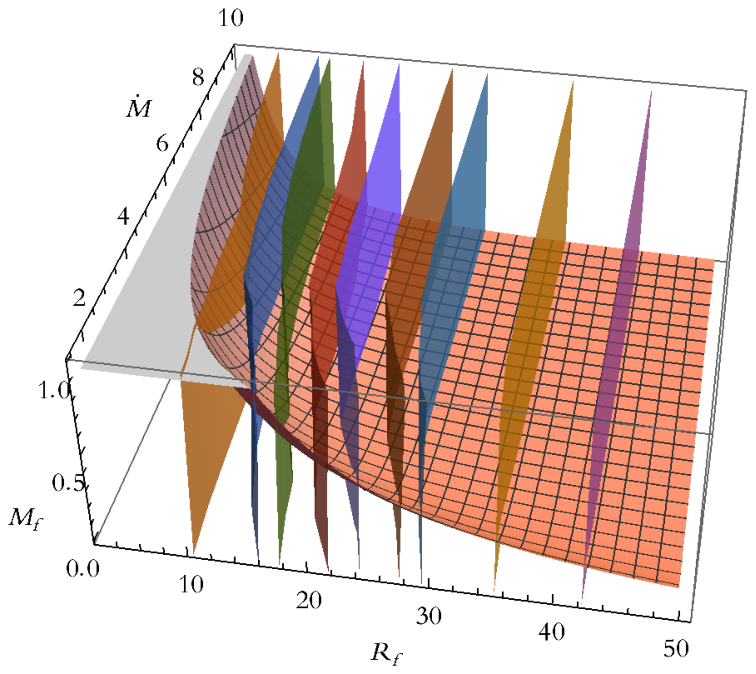}
 	 	\includegraphics[width=8.mm]{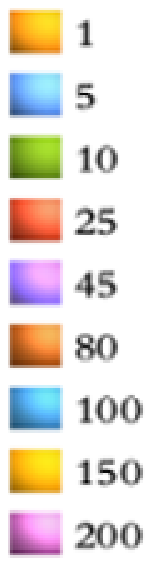}
 	 	
 	 	\includegraphics[width=80.mm]{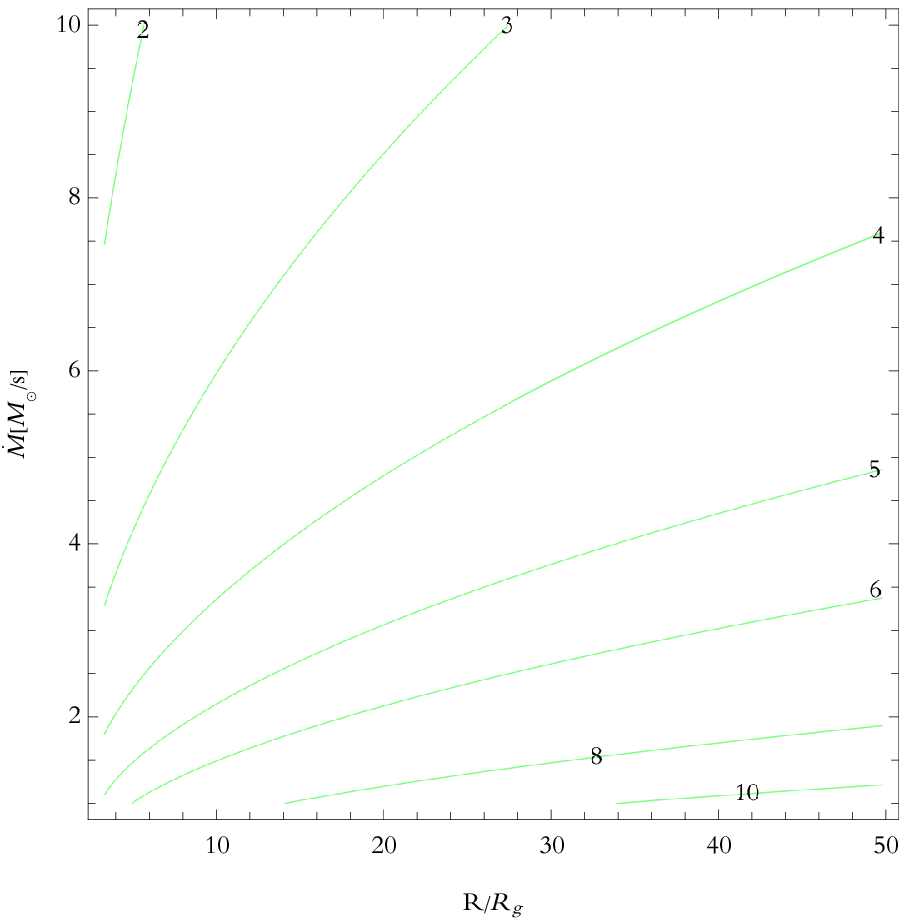}

 	\caption{\small{The top: The 3D panel illustrates the fragment mass $M_{f}~[M_{\odot}]$ versus $\dot{M}~[M_{\odot}~\rm s^{-1}]$, and radius $R_{f}$ at which the fragment is considered to be formed (this is actually an arbitrary value of $R/R_{g}$), shown as a pink surface. The colorful contours are associated with the delay time $t_{d}$, with their time values given in the plot legend. The bottom: This plot demonstrates the contours of $R_{\rm acc}$ in the $\dot{M}-R$ plane. }}
 	\label{fig3}
 \end{figure}

We found that fragmentation is possible in this unstable realm, for the cases with $\dot{M}\geqslant 1.3~M_{\odot}~\rm s^{-1}$, regarding $M_{\rm BH}=3~\rm M_{\odot}$ and $\beta=2.5\times 10^{-5}$ (see the plot in the bottom of Figure (\ref{fig2}) for more precise evaluation).

\subsection{Fragment accretion and observable evaluation}\label{sec3.1}

In order for a fragment to start accretion due to central BH's gravitational torque, as well as disk's viscous torque, the following condition should be met by the shear force per unit length $f_{\nu}$, tidal force $f_{T}$, and self-gravity force of the fragment $f_{SG}$,
\begin{equation}
f_{\nu}+f_{T}>f_{SG}.
\label{13}
\end{equation}
In other words, once the viscous and tidal torques (which are due to the shear force and that of the central BH's gravity) dominate over the torque associated with self-gravity of the fragment, viscous spreading of the clump is expected. \cite{2017MNRAS..464..4399} and \cite{2020ApJ..888..64} have adopted a similar approach for the same purpose as here.

Considering the fragment as a uniform spherical mass distribution, one may estimate $f_{T}=\frac{GM_{\rm BH}}{R^{2}}(\frac{l_{f}}{R})$, $f_{SG}=\frac{3GM_{f}}{5l_{f}^{2}}$, and $f_{\nu}=\frac{l_{f}\nu \Sigma R}{M_{f}}\mid \frac{\mathrm{d}\Omega}{\mathrm{d}R}\mid$, in which $M_{f}$ and $l_{f}$ are considered to be the fragment's mass and size. The latter is assumed as the local Jeans length, i.e. $l_{f}=\lambda_{J}=c_{s}\sqrt{\frac{\pi}{G\rho}}$. We also take the Jeans mass to approximate the  mass of the fragment, namely $M_{f}\approx M_{J}=\frac{4}{3}\pi \rho (\frac{\lambda_{J}}{2})^{3}$ with $\rho\approx \Sigma /2H$.

Through making use of the condition (\ref{13}), one can obtain the radius at which the accretion of each fragment is triggered ($R_{\rm acc}$). Having such a radius in hand, the viscous time during which a fragment is considered to be accreted, is achievable via Equation (\ref{11}).

To have a vision of the fragments' physical features, such as formation radius (an arbitrary value between $3R_{g}$ and the outer boundary of the unstable region $R_{f}$ (the blue solid line in the bottom panel of Figure (\ref{fig1})), fragment's mass $M_{f}$, the start point of the fragment's accretion $R_{\rm acc}$, and the time it takes for a fragment to migrate prior to accretion (delay time) $t_{d}$, we illustrate our results in Figure (\ref{fig3}), regarding $M_{BH}=3~\rm M_{\odot}$ and, in agreement with its proposed values \citep{2006ApJ...653...L89}, $\beta =2.5\times 10^{-5}$. As pointed out earlier, the migration can be considered as a result of two torques, one exerted by the viscous force and the other one is owing to gravitational interaction between central BH and fragment. The former brings about a migration time of the order of viscous time scale, $t_{\nu}$. For the latter, as discussed by \citet{1964Phys...Rev...136..1224}, the gravitational radiation leads such systems to encounter a secular change in their orbital angular momentum, which causes inspiral motion on a time scale \citep{1964Phys...Rev...136..1224,2007ApJ..658..1173}
\begin{equation}
t_{g}=\frac{5}{64\Omega}(\frac{GM_{f}^{3/5}M_{BH}^{2/5}}{c^{3}}\Omega)^{-5/3}.
\label{14}
\end{equation}
In order to estimate $t_{d}$, we assume the minimum value between these two time scales, i.e. $t_{\nu}$ and $t_{g}$, that reflects the dominant torque in making such a migration.

As mentioned previously, Figure (\ref{fig3}) includes representations of the fragments' features. The pink surface in the top panel depicts the fragments' mass behavior versus mass accretion rate and radius (we denoted the location of each fragment as $R_{f}$ which is actually an arbitrary value of the dimensionless radius $R/R_{g}$). The colorful 3D contours are associated with different $t_{d}~(s)$ with their time values given in the plot's legend. The plot in the bottom, on the other hand, demonstrates the contours of $R_{\rm acc}$ in the $\dot{M}-R$ plane. One can see the larger the accretion rate ($\dot{M}$) becomes the smaller the fragment's mass ($M_{f}$), and the accretion radius ($R_{\rm acc}$) get, while the delay time ($t_{d}$) increases. The decline in $M_{f}$ can be accounted for by the growth in density $\rho$, taking our results for $\Sigma$ and $H$ into consideration, as $\dot{M}$ rises. Although it is not possible to find a clear relation between $\dot{M}$ and $\rho$, it appears from outcomes that an increase in $\dot{M}$ leads to a larger growth in density comparing their relative variations, i.e. $\frac{\delta \rho}{\rho}> \frac{\delta \dot{M}}{\dot{M}}$. The thing that results in a larger increase in the self-gravity force with respect to the shear force. This, consequently, causes the fragment to migrate inwards a larger distance, through a longer time $t_{d}$, prior to accretion. Moreover, at a given accretion rate, fragments with smaller masses are created further in the disk. In other words, there is an inverse relation between $R_{f}$ and $M_{f}$. It gets quite obvious when one infers $M_{f}$ is proportional to $\frac{1}{\sqrt{M_{\rm disk}(R_{f})}}$, considering the previously introduced relations for $M_{f}$ and $l_{f}$.

Now we are at the point to estimate the consequent gamma-ray luminosity owing to the fragments' accretion. To this end, we take the following relation into account

\begin{equation}
L_{\gamma}=f\eta \dot{M}_{f} c^{2}
\label{15}
\end{equation}
where $\eta$ is the energy conversion efficiency from the rest-mass energy to the gravitational one which is considered to be $\sim 0.5$ \citep{2005Science..307..77}, and $f$ is taken as the conversion factor from the gravitational energy to the radiative one with a value of $\sim 0.01$ \citep{2007ApJ...663..437}. $\dot{M}_{f}$ can be approximated as $\sim M_{f}/t_{\nu}$.

\begin{figure}
 	\centering
 \includegraphics[width=89.5mm]{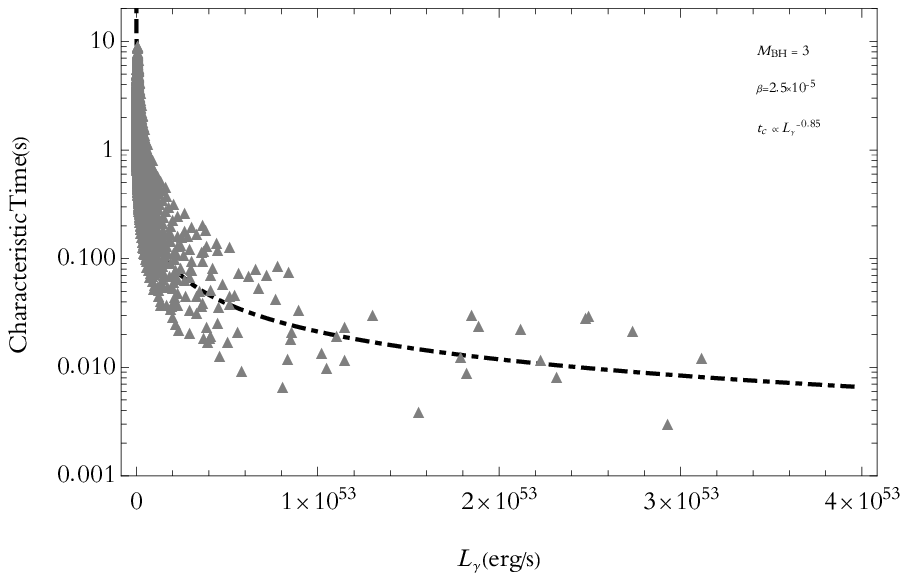}
 \includegraphics[width=89.5mm]{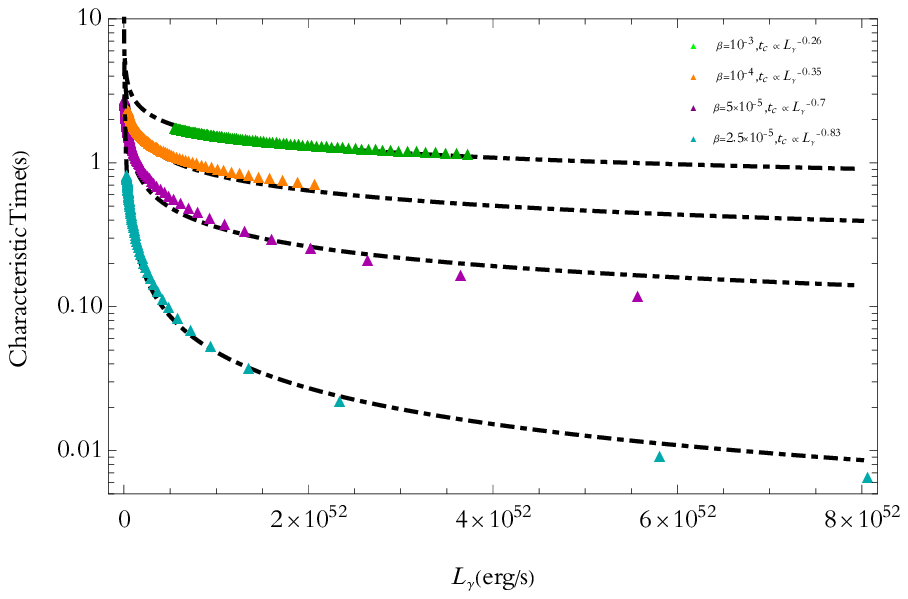}
 \includegraphics[width=89.5mm]{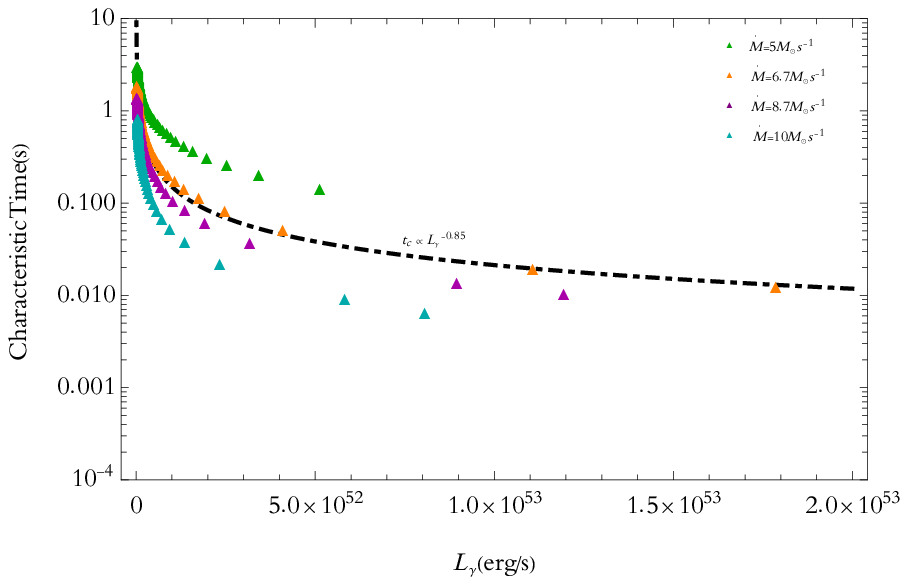}
 	\caption{\small{The top: The characteristic time behavior in terms of gamma-ray luminosity for a set of given parameters written inside the plot. Different mass accretion rates, range from $1.5-10~M_{\odot}~\rm s^{-1}$ are considered as well. Additionally, a variety of possible fragments that might form at different radii are taken into consideration. The black dot-dashed curve is the best fit line to our data.The middle: Different fits for various quantities of $\beta$ parameter and $\dot{M}=10~M_{\odot}~\rm s^{-1}$, are illustrated. Evidentally, as $\beta$ decreases, the fit to our data match better with the phenomenological findings. The bottom: $t_{c}$ versus $L_{\gamma}$, with a set of various mass accretion rates, i.e. $\dot{M}=5,6.7,8.7,$ and $10~M_{\odot}~\rm s^{-1}$ (with the curves in green, orange, purple, and cyan, respectively), and $\beta=2.5\times 10^{-5}$ are taken into account.}}
 	\label{fig4}
 \end{figure}


To provide an intuition about the validity of our model, we draw an analogy with empirical outcomes. There seem to be an anti-correlation between gamma-ray luminosity and the minimum variability timescale (regarded as a characteristic timescale $t_{c}$), namely $t_{c}\propto L_{\gamma}^{-1.0\pm 0.1}$ \citep{2015ApJ..805..86}. The latter is assumed to be the minimum timescale that separates white noise from red noise in a power density spectrum. There might be some alternatives to account for this anti-correlation in the literature, e.g. \citet{2015MNRAS..447..L11} argue that the interactions between rotating BHs with the surrounding matter can increase inversely proportional to duty cycle in advection of the magnetic flux. In our explanation of such a behavior, however, we approximate this characteristic timescale ($t_{c}$) to be the same as the viscous time of a fragment's accretion, $t_{\nu}$, and probe for its inverse correlation with the luminosity provided by each fragment's accretion.

We presented an illustration of the characteristic time behavior in terms of gamma-ray luminosity, in the top panel of Figure (\ref{fig4}). $M_{\rm BH}$ and $\beta$ are assumed to be $3~M_{\odot}$ and $2.5\times 10^{-5}$, with different mass accretion rates, range from $1.5-10~M_{\odot}~\rm s^{-1}$ are considered. It should be mentioned that we restricted ourselves to the area of the disk where our formalism is approximately valid (see section \ref{sec2}). Additionally, since it is of a remarkable uncertainty which fragment will be created throughout this region, we considered a variety of possible fragments that probably form at different radii. Regarding the best fit line (dot-dashed curve), a power law correlation between these two quantities can be confirmed, with a power index of about $-0.85$. This is interestingly close to its phenomenological value, i.e. $\sim -1$.

However, to find how our choices of $\beta$ parameter might affect the outcomes, we plotted different fits taking various quantities of this parameter into account, and considering $\dot{M}=10~M_{\odot}~\rm s^{-1}$ (the middle plot in Figure (\ref{fig4})). Obviously, the smaller the $\beta$ parameter gets the better our results satisfy the observations, so that it looks quite well-matched to the empirical results when $\beta\sim 10^{-5}$, the quantity which is very close to the value estimated by \citet{2006ApJ...653...L89}, for a self-gravitating unstable disk.

A similar analogy is provided for different mass accretion rates in the plot at the bottom of Figure (\ref{fig4}). $\dot{M}$ is assumed to be $5,~6.7,~8.7,$ and $10~ M_{\odot}~\rm s^{-1}$ with the curves in green, orange, purple, and cyan, respectively, and $\beta=2.5\times 10^{-5}$. Evidently, a growth in accretion rate gives birth to a decline in the characteristic timescale, which indicates a shorter temporal variability in the spectrum, that might be an expected consequence of the increase in $\dot{M}$.

Note that in the above analysis we assumed each fragment accretes due to $\beta$-viscosity. However, such an assumption might not be a good estimate since it appears through fragmentation there can be a transition from the so-called gravito-turbulent behavior (the turbulence due to gravitational instability) in the disk, which means this state might not be sustained after fragmentation (\cite{2016ApJ..824..91}). \{Therefore, regarding this uncertainty we consider other possible scenarios of the viscosity prescriptions that might be present during the fragment's accretion. We think that after fragmentation the density of material in the disk has been reduced because of fragmentation, so that the initial high neutrino opacity might not be reliable anymore (remember $\tau_{\rm tot} \propto \rho$). Hence, the MRI-turbulent accretion process seems to be a probable alternative. In this case, if the viscose time scale is shorter than the fallback time of the fragment to the pericenter radius (comparable to the tidal radius which has been denoted by $R_{\rm acc}$ and archived by equation (\ref{13})), one can approximate the accretion rate by the fallback rate \citep[for more details on this issue see e.g.][]{1990ApJ..351..38,2019ApJ..872..151}, so that the fallback time can be considered as $t_{c}$. Otherwise, the accretion time, $t_{\nu}$, would be another possible estimate for $t_{c}$. Considering $\alpha \sim 0.3$ \citep{2007MNRAS..376..1740,2019NewA..70..7} we found the former as the case in our formalism. Therefore, we assume the fall back time $t_{fb} \simeq 2\pi R_{\rm acc}^{3} \frac{1}{l_{f}^{3/2}\sqrt{GM_{\rm BH}}}$ \citep{Falanga2015,2020ApJL..896L..38C}, as the characteristic time scale of our interest. Such a consideration will change the power index of the best fit line, in the $L_{\gamma}-t_{c}$ panel, to $\sim -0.73$ with $t_{c}$ reaches an order of $\sim 0.005-0.05~\rm s$. Lower choices for $\alpha$ result in the longer minimum variability time scales, without making any significant change in the power index, which means a good agreement with observations.

Another issue that sounds worth noting is the probable mechanisms can result in the gamma-ray emission through the jet activity. As we claimed, the variability in the accretion rate can provide us with an inhomogeneous jet structure comprises the shells of different Lorentz factors. Regarding the variability time scale, which is mostly very short especially for the MRI-driven viscosity scenario (note that in the $\beta -\nu$ viscosity case, Figure (\ref{fig4}), the longer accretion times, corresponding to the outer fragments, might be due to an overestimation in the size of the neutrino trapped region), the jitter radiation \citep{2009ApJ..702..L91,2000ApJ..540..704} or the turbulent model \citep{2009MNRAS.. 394.. L117} can account for the mechanisms to produce the resultant spectrum. However, for the longer variability time scales which are attributed to the lower fragments' accretion rates (consider the lower clump's mass with a longer accretion time), it appears the large scale magnetic fields within the jet, play a dominant role in the emission mechanism via synchrotron radiation or Compton scattering.

\section{Fragments' Migration and GWs}\label{sec4}

As pointed out previously, prior to accretion, gravitational radiation and disk's dissipation effects might be responsible for the migration of the fragments, and consequently, a GW signature. To study its detectability, we need to make an estimate of the characteristic strain, $h_{c}$, which is identified as a dimensionless quantity that refers to the GW amplitude. To this end, imagine a system composed of a single fragment in a circular orbit around the BH. The GW strain amplitude for a source at the distance of $D=100D_{100}~\rm Mpc$ approximately reads the following relation \citep{2007ApJ..658..1173}

\begin{equation}
h_{0}=6.4\times 10^{-24} \Theta M_{f} (\frac{M_{BH}^{2/5}}{M_{\odot}})^{5/3}f_{100}^{2/3}D_{100}^{-1}
\label{16}
\end{equation}
where $\Theta=0-4$ is introduced in terms of orientation of the source and the antenna pattern of the detector \citep{2007ApJ..658..1173,1993Phys...Rev...D..47..2198}, and $f=100f_{100}~\rm Hz$ is the frequency. Long GRBs seem to be narrowly beamed in a normal direction to the accretion disk. Therefore, following \citet{2007ApJ..658..1173}, we consider our binary system to be of a low inclination that leads to $\Theta\approx 2.5$ \citep[see e.g.,][ for more details]{2007ApJ..658..1173,1993ApJ..417..L17}. The characteristic strain is then introduced as
\begin{equation}
h_{c}=h_{0}n_{cyc}^{1/2}
\label{17}
\end{equation}
where $n_{cyc}\equiv f^{2}/\dot{f}$ is the number of cycles spent within a bandwidth $\Delta f\sim f$ centered on $f$. Regarding $f=\Omega/\pi$ and taking the Keplerian angular velocity $\Omega =\sqrt{\frac{GM_{BH}}{R^{3}}}$ into account, this can be rewritten as $n_{cyc}=2fR/3\dot{R}$, with $\dot{R}/R$ is the total inward migration rate. One can approximate this factor as $|{\dot{R}/R}|=t_{\nu}^{-1}+t_{g}^{-1}$ \citep{2007ApJ..658..1173}.

\begin{figure}
 	\centering
 \includegraphics[width=83.mm]{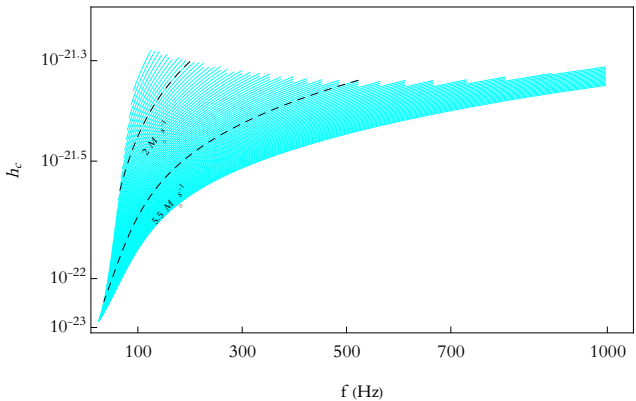}
 	\caption{\small{The characteristic strain in terms of frequency for a source distance of $100~\rm Mpc$. The curves in cyan are associated with different mass accretion rates, range from $1.5-10~\rm M_{\odot}s^{-1}$, the BH mass $3~ M_{\odot}$ and $\beta=2.5\times10^{-5}$. We distinguished two curves with $\dot{M}=2$ and $5.5~M_{\odot}~\rm s^{-1}$, through black color.}}
 	\label{fig5}
\end{figure}

\begin{figure}
 	\centering
 \includegraphics[width=88.mm]{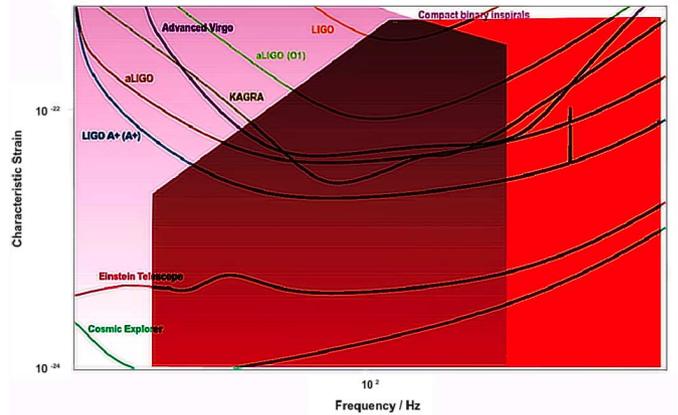}
 	\caption{\small{The noise amplitudes $h_{n}$ for different detectors are depicted through
colored curves. The area in pink demonstrates our model estimate of strain amplitude $h_{c}$ for a source at a distance $\geqslant 100~\rm Mpc$,
concerning $M_{\rm BH} = 3~M_{\odot}$ and $\beta = 2.5\times 10^{-5}$. For the sake of comparison with promising cases of astrophysical objects, data for compact binary inspirals is represented in bright purple. }}
 	\label{fig6}
 \end{figure}

In Figure (\ref{fig5}) we provided the outcomes for the characteristic strain in terms of frequency in which we assumed a source distance of $100~\rm Mpc$. The curves in cyan are associated with different mass accretion rates, range from $1.5-10~M_{\odot}~\rm s^{-1}$, the BH mass $3~M_{\odot}$ and $\beta =2.5\times 10^{-5}$. To provide a clear vision of $h_{c}$ trend versus frequency, we distinguished two curves with $\dot{M}=2$ and $5.5~M_{\odot}~\rm s^{-1}$, through coloring in black. For one thing, the higher mass accretion rates lead to the lower characteristic strain amplitudes in a wider range of frequency. The former reflects the fact that, for higher $\dot{M}$, although the number of cycles within a bandwidth increases due to the longer distance of inwards migration, the lower mass of the fragment at higher accretion rates gives birth to such a decline in $h_{c}$. For another, at a given accretion rate, as we go inward through the disk it appears the increasing fragment's mass plays a dominate role over the decreasing migration time of the fragments, and consequently the declining $n_{cyc}$, in order to produce GWs with higher characteristic strain.

More importantly, such GWs with $10^{-23}\lesssim h_{c}\lesssim 5\times 10^{-22}$ and $25~{\rm Hz}\lesssim f\lesssim 1000~\rm Hz$, are probably detectable via various GW detectors from LIGO (for the lower mass accretion rates) to Einstein Telescope (ET) and Cosmic Explorer (CE) (for the outer disk's fragments with higher $\dot{M}$). This detectability has been illustrated in Figure (\ref{fig6}), in which the sensitivity lines (the noise amplitudes $h_{n}$) of different detectors have been depicted through solid colored curves. The area in pink demonstrates our model estimate of $h_{c}$ for a source at a distance of $\geqslant 100~\rm Mpc$, considering $M_{\rm BH}=3~ M_{\odot}$ and $\beta =2.5\times 10^{-5}$. To compare with the other sources, we also demonstrated the realm of GW characteristic strains produced by compact binary inspirals that is represented in purple. This plot has been provided through the website \url{http://rhcole.com/apps/GWplotter}, in which similar figures can be generated \citep{2014CQG..32..015014}.

\section{Conclusions and discussion}
\label{sec5}

The main idea of the present work emanates from different approach adopted by \cite{2007ApJ...663..437} towards the neutrino function in the inner regions of NDAF, where neutrino opacity appears to be high enough to make neutrinos captured \citep{2002ApJ.. 579.. 706}. Such trapped neutrinos are considered as the main responsible for the energy and momentum transport \citep{2007ApJ...663..437}. The implementation of this idea in a self-gravitating NDAF, as what we conducted through this study, can account for the rapid fluctuations in GRBs' prompt emission, if the inner regions become gravitationally unstable and fragmentation gets probable. This is not a possible scenario, however, in case of a self-gravitating NDAF in which (magnetic) $\alpha$-viscosity is taken into consideration \citep[based on][for instance, only the outer disk is unstable when it comes to self-gravity in the presence of $\alpha$-viscosity]{2014ApJ...791..69,2017ApJ...845..64}. Our results indicate that such a formalism leads to the gravitational instability in the inner neutrino opaque regions, while the fragmentation does not occur. Hence, being inspired by the idea proposed by \citet{2006ApJ...653...L89}, we considered this instability as a source of turbulence and viscosity. To elaborate more, \citet{2006ApJ...653...L89} found that the functional form of this gravitationally driven viscosity is the same as that of the $\beta$ parameterizations, although they are based on different physical considerations. Through adding such a term to viscosity, we found it possible for the inner disk to undergo fragmentation. We studied the fragments' features to see if this clumpy structure gives birth to the temporal variability of the prompt emissions. The comparison of outcomes with observational findings showed that this scenario is successful, especially in the case of lower values for $\beta$ parameter ($\sim 10^{-5}$). This is in a good agreement with what has been estimated by \citet{2006ApJ...653...L89} for a self-gravitating unstable disk, on the other hand. Although, we emphasize that the uncertainties in adopting an assured strategy to trace the evolution of the clumps, need more precise studies through simulations to be treated. More interestingly, our simple model could predict the creation of GWs as a result of the fragments' migration prior to their accretion. The frequency band, $25~{\rm Hz}\lesssim f\lesssim 1000~\rm Hz$, and the amplitude of the waves, with $10^{-23}\lesssim h_{c}\lesssim 5\times 10^{-22}$, made it detectable for a range of current and future detectors from LIGO to CE.

There are some points worth discussing here. For one thing, in case of LGRBs, the initial higher accretion rates, i.e. $\dot{M}\gtrsim 1~M_{\odot}~\rm s^{-1}$ that appear to play a crucial role in our scenario to cause the fragmentation, are probable for the cases of the higher progenitor mass \citep[$\gtrsim 40-60~M_{\odot}$, see e.g.,][]{2019ApJ..878..142}. For the lower progenitor mass which is of a smaller initial mass accretion rate, however, the gravitational instability might result in a spiral wave structure that can be another source of variability in the accretion rate of materials towards the BH, as discussed by \cite{2007ApJ...663..437}. In case of SGRBs, on the other hand, one should note that the more massive fragments are not probable to be formed in the disk of a mass around $\sim 0.2-0.5 ~M_{\odot}$ (see, e.g., \cite{2015ApJ.. 815.. 54}). Moreover, regarding the accretion time of each clump (see Figure (\ref{fig4})), the outer fragments for the lower mass accretion rates ($\lesssim 5~M_{\odot}~\rm s^{-1}$) should also be ignored due to the accretion time longer than $~2~\rm s$ (the common time scale of the SGRB's prompt emission). Although, these considerations do not affect the final results remarkably.

For another, there are some assumptions we made in order to simplify our model, such as considering the gas pressure as the only effective factor in the total pressure. Granted such a consideration might be a rather good one, as pointed out by \citet{2002ApJ.. 579.. 706}, the effectiveness of the radiation pressure factor is probably worth investigating in such an advection dominated environment with trapped photons.

To shed light on the disk's cooling, we should add the consideration of the advection cooling, as the only process that cools down the disk, can be modified through taking the photodisintegration mechanism into account, as well. Other cooling processes, i.e. neutrino-antineutrino annihilation and radiation cooling, are not actually important in the inner disk due to the high neutrino opacity and the optically thick nature of this region. However, photodisintegration can affect the energy transpose of the system. As \citet{2007ApJ..658..1173} pointed out, this mechanism can be effective in cooling the outermost region of collapsar disks. The thing that can facilitate the fragmentation of the disk, in the case of being gravitationally unstable, via lowering the local cooling time scale. In our study we ignore this mechanism and considered the advection cooling process as the only one, for the sake of simplification. We think, taking the effects of photodisintegration into account can increase the fragmentation possibility, so that the main idea of the model will not be affected, although, the estimates we made for the minimum variability time scale and the strain amplitude may be varied due to a change in the fragments' features.

Finally, we found that the lower the $\beta$ parameter is the better the outcomes are confirmed by observations. However, bearing in mind $\beta$ is assumed to be proportional to the inverse Reynolds number \citep{2000A&A..357..1123}, this means a large Reynolds number is required to achieve better results. On the other hand, \citet{2007ApJ...663..437} pointed out the MRI can be effectively suppressed by the neutrino viscosity in case of small Reynolds numbers ($Re\lesssim 1$). This reflects the fact that for a self-gravitating NDAF such an assumption might not work well, and it should be investigated if the MRI can be suppressed by neutrino viscosity anymore. The thing that makes it of a probably great significance to consider the $\alpha$-prescription of viscosity, as well.

\section{acknowledgments}
This work was supported by the National Natural Science Foundation of China under grants 11822304 and 12173031. We hereby acknowledge Sci-HPC center of Ferowsi University of Mashhad where some part of this research was performed there.

\section{Data availability}
No new data were generated or analyzed in support of this research.

\bsp	
\label{lastpage}
\end{document}